\documentclass[10pt,letterpaper]{article}
\usepackage{opex3}
\usepackage{graphicx}

\begin{document}

\title{Enhancing image contrast using coherent states and photon number resolving detectors}

\author{A. J. Pearlman,$^{1,2}$ A. Ling,$^{1,2}$ E. A. Goldschmidt,$^{1,2}$ C. F. Wildfeuer,$^3$ \\J. Fan,$^{1,2}$ and A. Migdall,$^{1,2}$}
\address{$^1$Joint Quantum Institute, University of Maryland, College Park, MD 20742, USA\\
$^2$Optical Technology Division, National Institute of Standards and Technology, \\Gaithersburg, MD 20899, USA\\$^3$Hearne Institute for Theoretical Physics, Department of Physics and Astronomy,
\\Louisiana State University, Baton Rouge, LA 70803, USA}

\begin{abstract}We experimentally map the transverse profile of diffraction-limited beams using photon-number-resolving detectors. We observe strong compression of diffracted beam profiles for high detected photon number. This effect leads to higher contrast than a conventional irradiance profile between two Airy disk-beams separated by the Rayleigh criterion. 
\end{abstract}


\section{Introduction}
There is widespread interest in improving resolution and sensitivity in imaging and metrology applications \cite{SL08,TN07,JB96,NB00,MD01}. Many of the proposed improvements for precise phase measurements and high-resolution imaging employ highly nonclassical photon states such as photon-number (Fock) states and path-entangled photon-number (N00N) states due to their shorter de Broglie wavelength or squeezed states due to their suppressed noise \cite{NB00,JD08}. Unfortunately, such nonclassical states are highly sensitive to loss \cite{MR07}, and these schemes usually require the ability to resolve photon number efficiently \cite{CC81,RG08}, which has only recently become feasible in a research setting. This has led to interest in combining loss-tolerant coherent states with photon-number-resolving detectors, to realize some improvement over standard classical techniques and avoid the immense challenges associated with generating exotic quantum states \cite{SB04,CW09,GK06,PH06,NT03}.

In the context of imaging, much attention has focused on improving resolution beyond the Rayleigh limit \cite{LR79}. This limit is imposed by diffraction rather than by quantum fluctuations of the light \cite{NT03}. In particular, Giovannetti \emph{et al.} proposed that photon-number-resolving strategies could result in high-resolution images beyond the standard Rayleigh criterion \cite{VG09}. They note that using a coherent source and a photon-number-resolving measurement compresses the point spread function of an image but does not lead to improved image resolution. This compression could also be accomplished by classical post-processing of the data; however, it is not clear whether post-processing has an advantage in the presence of noise. Near the single-photon level, the presence of stray light degrades the utility of such post-processing, and classical detection itself becomes challenging. Nonetheless, direct detection of fringe compression can be used to improve the contrast between closely spaced diffracted beams. 

\section{Experimental Results}
\subsection{Setup}
Here, we report two experiments using photon-number-resolving detectors to study the spatial irradiance profile of diffracted laser beams. The main components of our experimental setups are a laser source, a photon number resolving detector and a single slit, or pinhole. We first diffract a Gaussian beam at a single slit and observe the expected narrowing of fringes with increasing detected photon number. We then use this fringe compression to demonstrate increased contrast of two beams with Airy disk profiles separated by less than the separation given in the Rayleigh criterion. 

Our photon-number-resolving detector, the transition edge sensor (TES), is a microbolometer capable of distinguishing photon number for a monochromatic source based on the amount of energy deposited on its superconducting tungsten film. The energy deposited causes a temporary change in current whose integral is proportional to the energy absorbed. Thus, output photoresponse pulses from the TES can be integrated to yield the number of photons absorbed in each pulse. Clear discrimination of photon number greater than ten has been observed for 1550 nm photons \cite{AM03}. Our 1550 nm laser diode source has linewidth $\approx0.1$ nm, is modulated at 50 kHz, and is pulsed with 100 ns-wide pulses. 
\subsection{Fringe Compression}
In our first experiment, we study the photon-number-resolved spatial profiles of a single-slit diffraction pattern. In the experimental setup, a beam exits a single-mode fiber and is collimated to generate a Gaussian profile with a beam waist several millimeters in diameter. It is then diffracted through a single $\approx$250 $\mu$m-wide slit. We scan a standard 9 $\mu$m core single mode fiber coupled to the TES detector across the diffraction profile in 50 $\mu$m steps, $\approx$23 cm from the slit, and detect a mean of $\approx$3.6 photons per pulse at the position of maximum irradiance. At each fiber position, we resolve photon number by integrating TES photoresponse pulses and placing them into histograms to reveal the photon number distribution \cite{AM03}. The detected photon numbers can be extracted from these histograms, and in this experiment, we distinguish up to nine photons (higher photon numbers are neglected due to low count rates). Thus, the spatial diffraction patterns of different photon-number states are extracted. 

\begin{figure}[htbp] \centering \includegraphics[width=10cm]{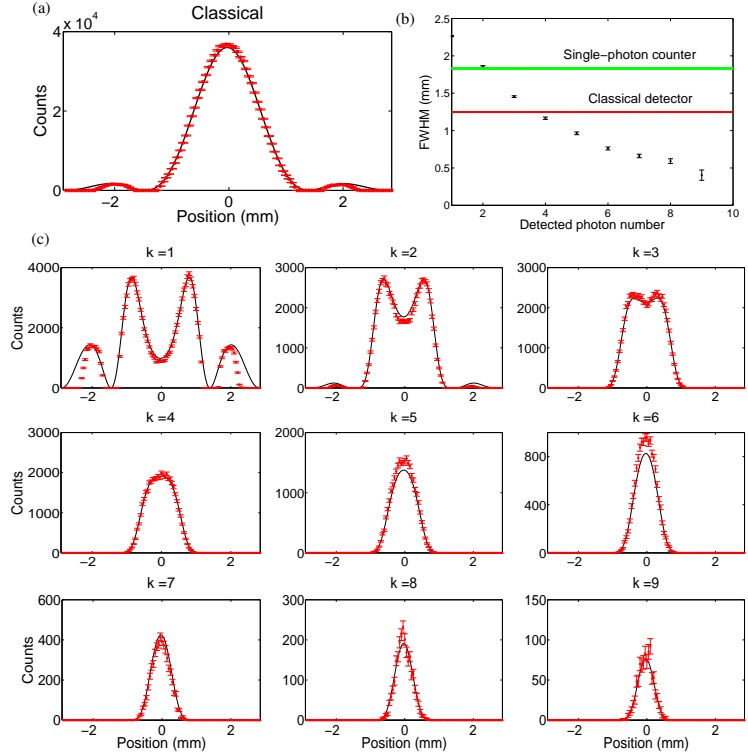} \caption{(a) Reconstructed classical spatial distribution with fit. (b) Full width at half maximum of the central fringe vs. photon number. Expected widths using reconstructed classical and conventional single-photon detectors shown for reference. (c) Spatial distribution of the diffraction pattern for up to nine photons with fits.} \end{figure}
  
We observe increased compression of the central lobe with increasing detected photon number, by up to a factor of three over a classical (average irradiance) signal (see Fig. 1). We reconstruct this classical signal from the photon-number-resolved measurements by taking $\sum kn_{k}$ where \emph{k} is the detected photon number and $n_{k}$ is the number of counts at a given \emph{k} (see Fig. 1a). We observe the $\rm sinc^{2}\left(\emph d\: \rm sin\theta/\lambda\right)$ dependence we expect from diffraction through a single slit, where \emph{d} is the slit width, $\lambda$ is the wavelength, and $\theta$ is the angle ($\theta\approx x_{t}/z$ where $x_{t}$ is the transverse position seen in Fig. 1 and $z\approx23$ cm is the distance from the diffracting slit to the detection plane). We then generate a fit to this curve by minimizing the least-squares error between $\mu\cdot\rm sinc^{2}\left(\emph d\: \rm sin\theta/\lambda\right)$ and the data by varying $\mu$ and $\theta$. We derive the photon-number profiles straightforwardly from the Poissonian distribution of a coherent source with a spatially varying detected mean photon number $\mu'=\mu\cdot\rm sinc^{2}\left(\emph d\: \rm sin\theta/\lambda\right)$ (see Fig. 1c).

This result can also be derived using an effective beamsplitter approach, as described in Ref. 11. In this case, the diffraction limited beam profile is modeled as a plane wave incident on a beamsplitter with spatially varying transmission coefficient \emph{T} \cite{CW09,GK06,CG05}. The equivalence between the two approaches can be shown directly in that the beamsplitter expression for the probability of detecting \emph{k} photons reduces to a Poisson distribution:
\begin{equation}
p(k)=  e^{-\mu}\sum_{j=k}^{\infty}\frac{\mu^{j}}{k!(j-k)!}\left|T\right|^{2k}(1-\left|T\right|^{2})^{j-k}
=e^{-\mu\left|T\right|^{2}}\frac{\left(\mu\left|T\right|^{2}\right)^{k}}{k!},
\end{equation}
where $\left|T\right|^{2}$ is replaced by the spatial profile of the beam, which in this case is the single slit far-field diffraction profile ($\rm sinc^{2}\left(\emph d\rm sin\theta/\lambda\right)$). 

\begin{figure}[htbp] \centering \includegraphics[width=12cm]{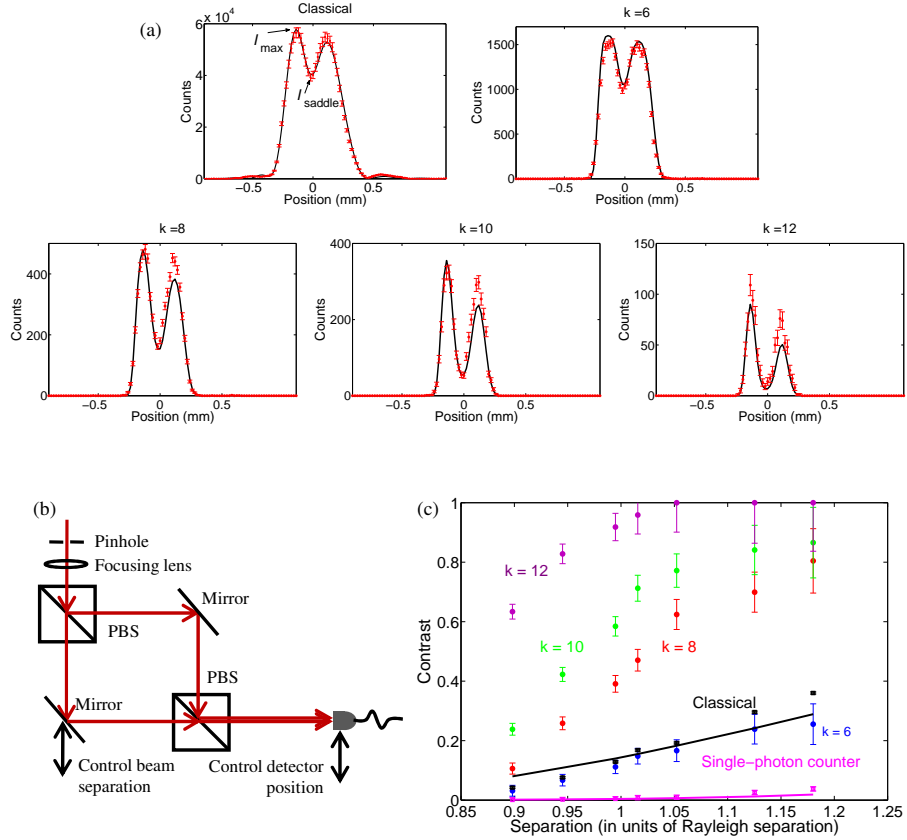} \caption{(a) Spatial profile at approximately the Rayleigh criterion for the reconstructed classical signal and photon-number detection at selected photon numbers larger than the mean photon number. (b) Experimental setup. PBS: polarizing beam splitter. (c) Contrast vs. beam separation (in units of the Rayleigh criterion) for selected photon numbers larger than the mean photon number. Contrasts derived from the reconstructed classical profile and single-photon counter profile are shown for reference.} \end{figure}
The compression of the central fringe FWHM we observe could also be obtained by post-processing of a classically detected signal.  However, because classical detection at low intensities is difficult, such measurements are often performed with conventional single-photon detectors (with no photon-number discrimination). We compare our results with what we would have obtained with a conventional single-photon detector, which records a photodetection event for every pulse with one or more detected photons. By treating the single and multi-photon detection events from all of the photon-number resolved profiles in Fig. 1(c) as identical photodetection events, we reconstruct the spatial profile obtainable with a conventional single-photon detector and find that our result is compressed by a factor of more than four (see Fig. 1(b)).
We note that the fringe width obtained with a conventional single photon detector is wider compared to a classical average irradiance measurement because the response of a non-number-resolving detector cannot scale with photon flux, but saturates with the detection of a single photon.
This illustrates the advantage of accessing the higher photon number statistics of the radiation field to yield higher compression over utilizing the average irradiance or conventional single-photon detection profile.

\subsection{Beam Contrast}
In the second experiment, we exploit the fringe compression observed above to study improved contrast of spatial profiles of two overlapping diffraction patterns with approximately equal irradiance, which are difficult to distinguish classically.  To obtain this profile, we first diffract a $\approx$2 mm diameter Gaussian profile beam through a standard 75 $\mu$m pinhole, obtaining a beam with an Airy disk profile. A 100 mm focal length lens, $\approx$155 mm after the pinhole, focuses the diffraction profile. The beam is split and recombined non-interferometrically using polarizing beam splitters in a Mach-Zehnder interferometer configuration (see Fig. 2b). We approximately equalize the photon flux in the arms using a fiber polarization controller and tune the spatial separation of the two beams by moving one of the mirrors. The detection system scans across the two nearly identical images for several different beam separation values while recording all photoresponse pulses. The maximum detected mean photon number per beam for all measurements is $\approx$5.3, and we distinguish photon numbers up to twelve. 

The Rayleigh criterion defines the classical limit for the minimum resolvable separation between two imaged, focused Airy disk profiles $\left(\frac{2J_{1}\left(\pi D\sqrt{x^{2}+y^{2}}/\lambda f\right)}{\pi D\sqrt{x^{2}+y^{2}}/\lambda f}\right)^{2}$, where $J_{1}$ is a Bessel function of the first kind, \emph{x} and \emph{y} define the position in the image plane, \emph{f} is the focal length, and \emph{D} is the aperture radius of the diffracting pinhole. This separation occurs where the main lobe of one beam falls on the first minimum of the other, and the angular separation with respect to the aperture is given by $1.22\lambda f/D$. The classical irradiance profile of two overlapping Airy disks is a saddle, and the contrast of this profile is defined as $C=\left(I_{\rm max} - I_{\rm saddle}\right)/\left(I_{\rm max} + I_{\rm saddle}\right)$, where $I_{\rm max}$ and $I_{\rm saddle}$ are the intensities at the peak and saddle points (see Fig. 2a).

The expected classical contrast for two equal intensity beams at the Rayleigh limit is  15~\%. We observe contrast values greater than 80~\% for detected photon numbers $k >> \mu$ at a separation of the Rayleigh limit (see Fig. 2c). At the smallest separation value studied, 90~\% of the Rayleigh limit, a contrast of over 60~\% is obtained for $k=12$. We note that the contrast value of 13~\% obtained from the reconstructed classical profile just below the Rayleigh limit matches closely with the theoretical Rayleigh limit value of 15~\%. As expected, the contrast of reconstructed profiles assuming the use of a conventional single photon detector is poor ($<5~\%$), because it does not exploit the full photon statistics. We show a sample set of measured photon-number-resolved profiles for a separation near the Rayleigh limit along with a theoretical fit derived analogously to the fit used for the single-slit configuration. Fitting parameters used account for the intensity difference of the two beams and the difference in optical path lengths. We show good agreement between the fits and data.
\section{Discussion}
We have thus far demonstrated an improved contrast between two Airy beam profiles at the Rayleigh criterion, using a coherent source which obeys Poissonian statistics. This raises the question of the effect of a source's statistical properties on the fringe compression and contrast measured with photon-number-resolving detectors. We point out that the compression of the central fringe is merely a consequence of the beam's photon statistics in combination with photon-number-resolved detection rather than a consequence of a particular aperture. Thus, this compression effect can be observed for any photon-number-resolved, diffraction-limited beam. To this end, we simulate photon-number-discriminated spatial distributions of Gaussian beam profile fields with coherent, thermal, and Fock statistics. We use the effective beamsplitter approach described above with a Gaussian profile replacing $\left|T\right|^{2}$ in Eq. 1 and the appropriate photon statistics. As shown in Fig. 3 for \emph{k} = 10, the Fock state ($\left|n\right\rangle = 10$) shows the highest degree of narrowing of the central fringe.

\begin{figure}[htbp] \centering \includegraphics[width=12cm]{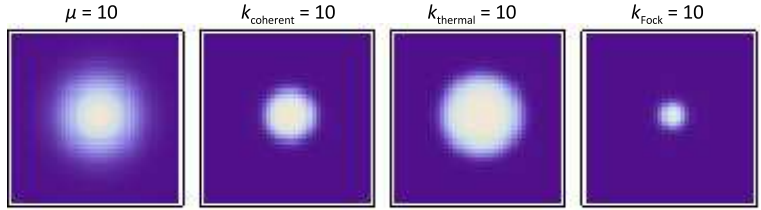} \caption{Calculated spatial distribution of (from left to right) a classically detected Gaussian profile coherent beam with a mean photon number of 10, a 10-photon detection of the same coherent state, a 10-photon detection of thermal light and same mean photon number, and a 10-photon detection of N=10 Fock state.} \end{figure}

This narrowing can translate directly into improved contrast if we add two coherent states or thermal states near the Rayleigh limit without interference as in our setup. The reason is that the compression from photon-number-resolved detection acts on the sum of the irradiance profiles of the beams \cite{VG09}. Thus, for these cases, the contrast is dictated completely by the amount of compression, which results in higher contrast for the Fock state profile over the thermal or coherent state profile. We note in this case that any increase in contrast can only occur for separations larger than the Sparrow limit, $\approx$ 0.8 of the Rayleigh criterion, where the two overlapping beam profiles combine to form a flat top \cite{CS16}. The reason is that the photon-number-resolving detection amplifies the dip feature in profiles between two beams. In the case of two beams separated by less than the Sparrow limit, no such feature can be exploited. Ref. 16 describes strategies employing ``incoherent mixtures'' of coherent states or Fock states in which compression from a photon-number-resolving detector acts on irradiance profiles individually.  Using such strategies, the Sparrow limit could potentially be surpassed.

As mentioned earlier, some fringe compression can be obtained via post-processing of a classical irradiance signal. Since this post-processing uses only the irradiance profile, it is not dependent on the photon statistics of the field and moving to a source with thermal, Fock, or other statistics has no effect. The amount of compression one can obtain via this method depends entirely on the signal to noise ratio of the detected signal, which is the reason it is difficult at low light levels. In addition, because classical post-processing can only amplify existing features in the profile, it can never resolve two beam profiles separated by less than the Sparrow limit.

In conclusion, we have experimentally observed the photon number resolved transverse profile of diffracted beams. We directly observe fringe compression of the diffraction pattern that is not possible with classical detectors or conventional single photon detectors. We further demonstrate that this fringe compression allows increased contrast of two nearly overlapping Airy disk profiles and discuss the effect of the source statistics on fringe compression. These studies may be useful for designing better metrology and imaging techniques. 
\section{Acknowledgements}
A.J.P. acknowledges support from the Intelligence Community Postdoctoral Research Associateship Program. We would like to thank Sergey Polyakov and Jun Chen for helpful discussions.
\end{document}